\documentclass[prl,twocolumn]{revtex4}
\usepackage{mathrsfs}
\usepackage{amssymb}
\usepackage{amsmath}
\usepackage{amsfonts}
\usepackage{graphicx}           
\usepackage{rotating}    
\usepackage{dcolumn}
\usepackage{epsfig} 

\begin{document}


  \title{Upper bounds in phase synchronous 
weak coherent chaotic attractors} 
\author{M. S.  Baptista, T. Pereira, J. Kurths}
\address{Universit{\"a}t Potsdam, Institut f{\"u}r Physik Am Neuen
    Palais 10, D-14469 Potsdam, Deutschland} \date{\today}

\begin{abstract}  
  An approach is presented for coupled chaotic systems with weak
  coherent motion, from which it is estimated the upper bound value
  for the absolute phase difference in phase synchronous states.  This
  approach shows that synchronicity in phase implies synchronicity in
  the time of events, characteristic explored to derive an equation to
  detect phase synchronization, based on the absolute difference
  between the time of these events.
\end{abstract}

\maketitle 

\section{Introduction}

This work deals with the phenomenon of phase synchronization (PS)
\cite{livro,park} in coupled chaotic systems, which describes
interacting systems that have a bounded phase difference, despite of
the fact that their amplitudes may be uncorrelated.  PS was found in
many natural and physical systems \cite{livro,park}, being
experimentally observed in electronic circuits \cite{parlitz}, in
electrochemical oscillators \cite{hudson}, in the Chua's circuit
\cite{baptista:2003}, and in spatio-temporal systems \cite{epa}. There
are also evidences of PS in communication processes in the Human brain
\cite{fell:2002,mormann:2003}.

In the case of two coupled systems, PS exists \cite{livro} if
\begin{equation}
|\phi_1 - q \phi_2| \leq \varrho,
\label{ps_cond_I}
\end{equation}
where $\phi_{1,2}$ are the phases calculated from a projection of the
attractor onto appropriate subspaces $\mathcal{X}_{1,2}$, in which the
trajectory has coherent properties \cite{livro,kreso}.  The rational constant
$q$ \cite{cita_r} is the frequency ratio between the average phase
growing and $\varrho$ is a finite constant to be determined, bounded away 
from zero.

The purpose of this work is to give an upper bound value for the
absolute phase difference in Eq. (\ref{ps_cond_I}) in phase
synchronous states, in terms of a defined phase
\cite{phase_definitions}. This is equivalent to determine an inferior
bound value for the constant $\varrho$. We show that this minimal
value, namely $\langle r \rangle$, can be estimated as the average
growing of the phase, calculated for typical trajectories, in one of
the subspaces.  Particularly, $\langle r \rangle$= $\langle W \rangle
\times \langle T \rangle$, where $\langle W \rangle$ is the average
angular frequency associated to a subspace $\mathcal{X}$, and $\langle
T \rangle$ is the average returning time of trajectories in this same
subspace, calculated from the recurrence of events of the chaotic
trajectory.  Similarly to periodic oscillating systems, in which it is
valid to say that an angular frequency $\omega$ is related to the
period $T$ by $\omega$ = 2$\pi/ T $, for chaotic systems it is valid
to say that $\langle W \rangle$ = $\langle r \rangle / \langle T
\rangle$.

In the derivation of the constant $\langle r \rangle$, we obtain a
series of inequalities that can be used to check for the existence of
PS.  A particular interesting one is suitable for systems where the
only available information is a series of time events.  We also
introduce the phase of a chaotic attractor to be given by the amount
of rotation of the tangent vector of the flow.

These results are shown to be valid to weak coherent attractors.  By
weak coherent attractors we mean following Ref. \cite{kreso},
attractors in which it is possible to define a Poincar\'e section or a
threshold that defines an event, such that for the time between two
events $\tau$, it is true that $| \tau - \langle T \rangle |<\kappa$,
where $\langle T \rangle$ is the average returning time between two
successive events, and $\kappa < \langle T \rangle$ is a small
constant.  So, our results are extended to attractors whose
trajectories might not have a clear rotation point, but still
presenting a weak coherent property in the time between events, e.g.
bursting/spiking dynamics.

For illustrating our ideas, we use two coupled R\"ossler oscillators,
and two coupled neuron models from the Rulkov map \cite{rulkov}. This
last example was chosen because we want to demonstrate that PS can be
detected by only knowing the time at which bursts occur (events).

\section{A minimal bound for the constant $\varrho$}

We start by developing some ideas to give a minimum bound for $c$ in
Eq.  (\ref{ps_cond_I}). For simplicity, we eliminate the rational
constant $q$ \cite{cita_r}, given by $q$=$\frac{\langle W_1
  \rangle}{\langle W_2 \rangle}$, by a changing of variable,
$\phi_2(t)^{\prime}$ = $q\phi_2(t)$.  With a slight abuse of notation,
from now on, we omit the $\prime$ symbol in the phase. Note however
that such a changing of variable does not change the fact that PS
exists or not.

Having two oscillators $S_1$ and $S_2$ that are coupled forming the
attractor $\Sigma$, we define the susbspaces $\mathcal{X}_{j}$ to be a
special projection in the variables of  $\Sigma$. This
projection is such that the attractor in this subspace presents the
coherent properties defined in \cite{kreso}. Subspaces
$\mathcal{X}_{j}$ are the same ones where the phase is calculated.

Next, we define a time series of events, where events here are 
the crossing of the trajectory to a given Poincar\'e section, some
local maxima/minima, or the crossing of one variable to some threshold.
Being $\Sigma$ the attractor of the coupled system, and
$\mathcal{X}_{1}$ and $\mathcal{X}_{2}$ two subspaces (on which the
phase is defined), $\tau_{j}^i$ is the time at which the $i$-th event
happens in $\mathcal{X}_{j}$. We consider the average return time,
$\langle T_j \rangle$, of the subspace $\mathcal{X}_j$ to be the
average of time intervals $T_j^i$=$\tau_j^{i}-\tau_j^{i-1}$ between
two events in $\mathcal{X}_j$, for $N$ events.  So,
\begin{equation}
\langle T_j \rangle = \frac{\sum_{i=1}^{N} T_j^i}{N}= \frac{\tau_j^N}{N}.
\label{media_tempo}
\end{equation}

We introduce the phase as the amount of rotation of the unitary tangent 
vector, $\vec{\mathcal{A}}_j(t)$. Being $|\vec{\mathcal{A}}_j(t+ \delta t) - 
\vec{\mathcal{A}}_j(t)|$ a small displacement of the phase for the time 
interval $\delta t$, calculated on the subspaces $\mathcal{X}_j$, and  
making $\delta t \rightarrow 0$, we arrive at
\begin{equation}
\phi_j(t)=\int_{0}^{t}|\dot{\vec{\mathcal{A}_j}}| dt
\label{integral_vetor_tangente}
\end{equation}
\noindent
So, $\phi_j(t)$ measures how much the tangent vector of the flow,
projected on the subspaces $\mathcal{X}_j$, rotates in time.  This
equation also suggest that $|\dot{\vec{\mathcal{A}_j}}|$ can be seen
as an angular frequency, more precisely $W_j = \left| \frac{ d
    \vec{\mathcal{A}_j}}{d t} \right|$ \cite{calcula_analitico}. 
 and the average angular frequency is simply
\begin{equation}
\langle W_j \rangle =\frac{1}{T} \int_0^{T} W_i dt.
\label{ergodic_02}
\end{equation}

We introduce the quantity $r_1^i$ =
$\int_{\tau^i_{2}}^{\tau^{i+1}_{2}} W_1 dt$,
which is the evolution of the phase from the time $\tau^i_2$ (when the
$i$-th event happens in $\mathcal{X}_2$) until the time $\tau^{i+1}_2$
(when the ($i+1$)-th event happens in $\mathcal{X}_2$). Thus, $\langle
r_1 \rangle = (\sum_{i=0}^{N} r_1^i)/N$, or in a continuous form,
after the $N$-th event, this average is calculated as 
$\langle r_1 \rangle = \sum_{i} \int_{\tau_2^i}^{\tau_2^{i+1}} W_j dt/N$ which
is equal to 
\begin{equation}
\langle r_1 \rangle = \frac{\int_0^{\tau_2^N} W_1 dt}{N}.
\label{media_r}
\end{equation}
\noindent
Using that it is valid to say that $\tau_2^N \approxeq N \langle T_1 \rangle$.
For $\tau_2^N \rightarrow \infty$, in Eq. (\ref{ergodic_02}) we have that
$\langle r_1 \rangle = \frac{1}{N \langle T_1 \rangle}\int_0^{\tau_2^N} W_1 dt 
= \frac{1}{\langle T_1 \rangle}\left(  \frac{1}{N}\int_0^{\tau_2^N} W_1 dt  \right)$, which
using Eq. (\ref{media_r}), can be written as
\begin{equation}
\langle W_1\rangle = \frac{\langle r_1 \rangle}{\langle T_1 \rangle}
\label{relaciona_r_T}
\end{equation}
\noindent
\noindent

These calculations can be done for $\langle r_2 \rangle$, however, if
PS exists, i.e. Eq.  (\ref{ps_cond_I}) is satisfied, one should have
that $\langle W_1 \rangle$=$\langle W_2 \rangle$, $\langle r_1
\rangle$=$\langle r_2 \rangle$, and $\langle T_1 \rangle$ = $\langle
T_2 \rangle$. Thus, in Eq.  (\ref{relaciona_r_T}) we can use the index
$j$.

{\bf Synchronicity of events: } The number of events at a given time
for synchronous oscillators is not always the same, but can differ by
an unity. This occurs because the $N$-th event in $\mathcal{X}_1$ and
$\mathcal{X}_2$ may not be simultaneous, resulting in a difference of
an unity between the number $N_1$ and $N_2$ of events, in
$\mathcal{X}_1$ and $\mathcal{X}_2$, respectively.  So, we can say
that the number of events in PS are always related by
\begin{equation}
| N_1(t) - N_2(t)| \leq 1.
\label{event_synchrony}
\end{equation}

The inequality in Eq. (\ref{event_synchrony}) is another variant for
an equation already used to detect phase synchronization \cite{livro}.
In that equation, every time an event occur, like the crossing of the
trajectory through a treshold, the phase is assumed to grow 2$\pi$.
And PS is considered to happen if the phase difference is always
smaller or equal than 2$\pi$. Note that Eq.  (\ref{event_synchrony})
can also be used to detect synchronous events in maps, in the case an
event can be well specified. As an example, one can observe the
occurrence of local maxima in the trajectory \cite{comment1}.

{\bf Synchronicity in the time of events}: 

Using Eq. (\ref{media_tempo}) in  
Eq. (\ref{event_synchrony}), we arrive at:

\begin{equation}
\left|\sum_{i=0}^{N} (T_1^i - T_2^i) \right| \leq \langle T_1 \rangle.  
\label{timediff}
\end{equation}
\noindent

This equation is related to the weak coherence in the dynamics.  The
more phase coherent the attractors are the more the amount $|\sum_i
(T_1^i - T_2^i) |$ approaches to zero.  As a consequence, the value
$\langle T_1 \rangle$ over estimate the maximum difference in the time
intervals between events. To overcome this, we introduce a physical
parameter, namely $\gamma$, which brings us information about the
coherence of a specific system. Thus, we put Eq. (\ref{timediff}) as

\begin{equation} | \sum_{i=0}^{N}
  (T^i_1 - T^i_2) |\leq \gamma \langle T_1 \rangle, 
\label{discrete_difference}
\end{equation}
\noindent
It is important to notice that $\gamma$ also
brings some information about the projection and about the section in
which the events are defined, once that the difference in the time intervals depends on the
projection and on the Poincar\'e section definition. Our calculations 
to the coupled R\"ossler-like attractors, show that $\gamma = 1/2$.

Multiplying both sides of Eq.  (\ref{discrete_difference}) by $\langle
W_1 \rangle$, we can relate time of events with the 
averaging growing of the phase:

\begin{equation}
\langle W_1 \rangle | \sum_{i=0}^{N} (T^i_1 - T^i_2)|
\leq  \gamma \langle r_1 \rangle.
\label{nova1}
\end{equation}

{\bf Synchronicity of the phase: }  
Next, we represent Eq. (\ref{ps_cond_I}) at the time the $N$-th event happens in 
$\mathcal{X}_1$, by 
\begin{equation}
\left| \sum_{i=0}^{N-1} (r^i_1 - r^i_2) + \xi(N) \right| \leq  \langle r \rangle, 
\label{nova2}
\end{equation}
\noindent
where
\begin{equation}
\xi(N)=\int_{\tau_1^{N-1}}^{\tau_1^{N}} W_1 dt -
\int_{\tau_2^{N-1}}^{\tau_1^{N}} W_2 dt 
\label{xi}
\end{equation}
The term $\sum_{i=0}^{N-1} (r^i_1 - r^i_2)$ represents the phase in
$\mathcal{X}_1$ at the moment the $(N-1)$-th event happens in
$\mathcal{X}_2$ minus the phase in $\mathcal{X}_2$ at the moment the
$(N-1)$-th event happens in $\mathcal{X}_1$. The term $\xi(N)$
represents the difference between the evolution of the phase from the
event $N-1$ till the time at which the $N$-th event happens at the
subspace $\mathcal{X}_1$, minus the evolution of the phase at the
subspace $\mathcal{X}_2$ from the ($N-1$)-th event in $\mathcal{X}_2$
until the time at which the event $N$ happens in $\mathcal{X}_1$.
This term establishes a bridge between the continuous-time formulation
of the phase difference [Eq. (\ref{ps_cond_I})] and the phase
difference between events.

From Eq. (\ref{nova1}), one sees that the smaller (bigger) the time
difference $| T_1^i - T_2^i|$ is the more (the less) synchronous the
system is, which means that the phase difference $| r_1^i - r_2^i|$
also gets smaller (bigger).  So, it is suggestive to consider that the
difference $(r^i_1 - r^i_2)$ is linearly related to $(T^i_1 - T^i_2)$
as
\begin{equation}
(r^i_1 - r^i_2) = \beta \langle W_1 \rangle (T^i_1 - T^i_2) + \sigma(i), 
\label{relacao_fase_tempo}
\end{equation}
\noindent
with $\beta$ being a constant, and $\sigma(i)$ brings the non-linear
terms.  

To obtain the value of the constant $\beta$, we imagine that PS is
about to be lost, by a small parameter change, and so, $|T^N_2 -
T^N_1|$ approaches $\gamma \langle T_1 \rangle$. Analogously, at this
situation, the phase difference $|r_1^i - r_2^i|$ has the ability to
grow one typical cycle, i.e., $\langle r_1 \rangle$, and
therefore, the term $\sigma(i)$ in Eq. (\ref{relacao_fase_tempo})
becomes very small and can be neglected. Thus, we have that $\beta
\langle W_1 \rangle \gamma \langle T_1 \rangle = \langle r_1 \rangle$,
and we arrive at $\beta = \frac{1}{\gamma}$.  For coherent attractors,
e.g.  R\"ossler-type, $\beta$ is approximately 2. This result is
discussed in the Appendixes. In Appendix \label{apendice1}, we discuss
how to construct maps using the time events $\tau_j^i$, and in
Appendix \label{apendice2}, we explain how to use these maps in order
to obtain that $\beta$=2.

Knowing the constant $\beta$, we put Eq. (\ref{relacao_fase_tempo}) in  
Eq. (\ref{nova2}), and we have that
\begin{equation}
|\beta \langle W_1 \rangle  \sum_{i=0}^{N-1}(T^i_1 - T^i_2) + 
\sum_{i=0}^{N-1} \sigma(i) + \xi(N)| \leq \varrho.
\label{relacao22}
\end{equation}
\noindent

Using the triangular inequality and the fact that $\varrho$ at the moment is considered to be 
an arbitrary constant, with a threshold (minimal) value, we write that
\begin{equation}
|\beta \langle W_1 \rangle  \sum_{i=0}^{N-1}(T^i_1 - T^i_2)| + 
|\sum_{i=0}^{N-1} \sigma(i) + \xi(N)| \leq \varrho.
\label{tiago1}
\end{equation}
\noindent
Equation (\ref{tiago1}) can be written as $|\sum_{i=0}^{N-1} \sigma(i)
+ \xi(N)| \leq \varrho - |\beta \langle W_1 \rangle
\sum_{i=0}^{N-1}(T^i_1 - T^i_2)|$.  At a specific event, may the
variable $|\sum_{i=0}^{N-1} \sigma(i) +\xi(N)|$ reaches the permitted
maximum value, this implies that the variable $|\beta \langle W_1
\rangle \sum_{i=0}^{N-1}(T^i_1 - T^i_2)|$ gets close to zero. At this
situation, $|\sum_{i=0}^{N-1} \sigma(i)| \leq \varrho$. Using the same
arguments we arrive at $|\beta \langle W_1 \rangle
\sum_{i=0}^{N-1}(T^i_1 - T^i_2)| \leq \varrho$, which implies that $
\langle r_1 \rangle \leq \varrho$.  Since $|\sum_{i=0}^{N-1} \sigma(i)
+max \xi(N)| \le \varrho$ we also have straithforward that
$|\sum_{i=0}^{N-1} \sigma(i)| \le \varrho$.

These results shows that the upper bound for the phase difference is
given by the constant $\langle r_1 \rangle = \langle W \rangle \times
\langle T \rangle$.  This means that the arbitrary constant $\varrho$
in Eq. (\ref{ps_cond_I}) is always greater than or equal to $\langle
r_1 \rangle$, in other words, $\langle r_1 \rangle$ is our threshold.
The physical meaning is obvious. If $\langle r_1 \rangle$ is the bound
for phase difference, given a number $\kappa \geq 1$, the value
$\kappa \langle r_1 \rangle$ is also a bound, but it is not a minimal
one.  Thus, we fix the constant $\varrho$ as
\begin{equation}
\varrho = \langle r_1 \rangle.
\label{valor_c}
\end{equation}
\noindent 
From Eqs.  (\ref{tiago1}), and (\ref{ps_cond_I}), 
we have the following inequalities
\begin{eqnarray}
 | \sum_{i=0}^{N-1} \sigma(i)| &\leq& \langle r_1 \rangle \label{novo_bound} \\
 \beta \langle W_1 \rangle  | \sum_{i=0}^{N-1}(T^i_1 - T^i_2) | &\leq & \langle r_1 \rangle 
\label{relacao1} \\
| \sum_{i=0}^{N-1} (r^i_1 - r^i_2)| &\leq& \langle r_1 \rangle. \label{nova3} \\
 |\phi_1(t) - \phi_2(t)| &\leq&  \langle r_1 \rangle, \label{PS}
\end{eqnarray}
\noindent

If one wants to use the inequality in Eq. (\ref{PS}) [or Eq.
(\ref{nova3})] to detect phase synchronization, it is required that
the phase is an available information.  For that, one needs to have
access to a continuous measuring of at least one variable. The
inconveniences of using this approach becomes evident when either one
has an experimental system where the only available information is a
time series of events, like the dripping faucet experiment
\cite{sarto} or the signal is so corrupted by noise that one can
really only measure spikes in neurons \cite{roland}. In these two
cases one should use the inequality in Eq.
(\ref{discrete_difference}) [or Eq. (\ref{relacao1})]. The only
inconvenience in the use of this inequality is that one should be
careful with the type of event chosen. If the specified event is the spiking
times, one might not see PS in the bursting time (and vice-versa). In
detecting PS in large networks, it might be computationally costy to
check for all the phase difference or event time difference among each
pair of subsystems. In this case, one could check the validation of
inequality in Eq.  (\ref{event_synchrony}), having in mind that such a
condition is only a necessary condition for PS.

\section{Phase synchronization in the two 
coupled R\"ossler oscillators}

\begin{figure}[!h]
\centerline{\hbox{\psfig{file=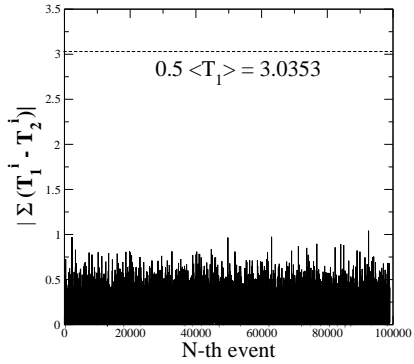,height=6cm}}} 
\caption{The fluctuation 
  $| \sum_{i=0}^{N} (T^i_1 - T^i_2) |$, in Eq.
  (\ref{discrete_difference}). Note that $| \sum_{i=0}^{N} (T^i_1 -
  T^i_2) |$ $\leq 0.5 \langle T_1 \rangle$. $\delta
  \alpha=0.001$ and $\epsilon$=0.01.  The phase is calculated from Eq.
  (\ref{integral_vetor_tangente}).}
\label{nova_fig20}
\end{figure}

To illustrate our approach we consider two non-identical 
coupled R\"ossler   oscillators, given by 
\begin{eqnarray}
\dot{x}_{1,2} &=& -\alpha_{1,2}y_{1,2}-z_{1,2}+\epsilon
[x_{2,1}-x_{1,2}] \nonumber \\ 
\dot{y}_{1,2} &=& \alpha_{1,2}x_{1,2}+ay_{1,2} \label{rossler} \\ 
\dot{z}_{1,2} &=& b+z_{1,2}(x_{1,2}-d) \nonumber,  
\end{eqnarray}
\noindent
with $\alpha_{1}=1$, and $\alpha_{2}=\alpha_{1}+\delta \alpha$. The
constants $a$=0.15, $b$=0.2, and $d$=10 are chosen such that we have a
chaotic attractor in a phase coherent regime. The subspace where the
phase is computed is given by $\mathcal{X}_1=({x_1,y_1})$ and
$\mathcal{X}_2=({x_2,y_2})$.  The time series, $\mathcal{M}^{\tau}_j$,
that define the events in $\mathcal{X}_j$, are defined as follows.
$\mathcal{M}^{\tau}_1$: {\bf $\tau^i_1$} is constructed measuring the
time the trajectory crosses the plane $y_2$=0 in {\bf
  $\mathcal{X}_2$}; $\mathcal{M}^{\tau}_2$: {\bf $\tau^i_2$} is
constructed measuring the time the trajectory crosses the plane
$y_1$=0 in {\bf $\mathcal{X}_1$}.  In Fig.  \ref{nova_fig20}, we show
the coupled R\"ossler oscillator for the parameters $\delta
\alpha=0.001$ and $\epsilon$=0.01. We show that Eq.
(\ref{discrete_difference}) is always satisfied (for $10^5$ pairs of
events), i.e., $| \sum_{i=0}^{N} (T^i_1 - T^i_2) |$ $\leq \langle T_1
\rangle /2 $, with $\langle T_1 \rangle /2 $=3.0353.

\begin{figure}[!h]
\centerline{\hbox{\psfig{file=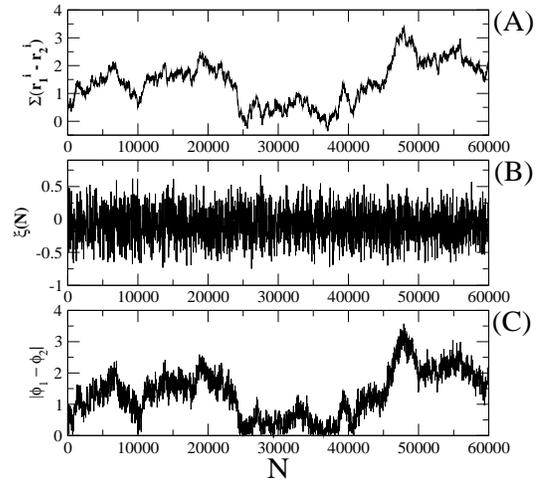,height=8cm,width=8cm}}} 
\caption{In (A), we show the phase difference at the time
  the $N$-th event happens in both subsystems. In (B) we show
  $\xi(N)$, and in (C), we show the phase difference, at the time that
  the $N$-th event happens in $\mathcal{X}_1$.  So, the number of
  events in $\mathcal{X}_2$, $N_2$, can assume either one of the
  following values $(N-1, N, N+1)$. $\delta \alpha=0.001$ and
  $\epsilon$=0.01.  The phase is calculated from Eq.
  (\ref{integral_vetor_tangente})}
\label{nova_fig7}
\end{figure}
In Fig. \ref{nova_fig7}(A), we show the phase difference at the time
the $N$-th event happens in both systems, i.e. the term $\sum_{i=0}^{N}
(r^i_1 - r^i_2)$ in Eq. (\ref{nova2}). Note that the time that the
$N$-th event happens in $\mathcal{X}_1$ is different that the time the
$N$-th event happens in $\mathcal{X}_2$.  In (B), we show $\xi(N)$ in Eq.
(\ref{xi}), and in (C), we show the phase difference, at the time that
the $N$-th event happens in $\mathcal{X}_1$.  Note that the phase
difference in (C) is just the phase difference for
the same number of events [in (A)] plus the term $\xi(N)$ [in (B)].

\begin{figure}[!h]
\centerline{\hbox{\psfig{file=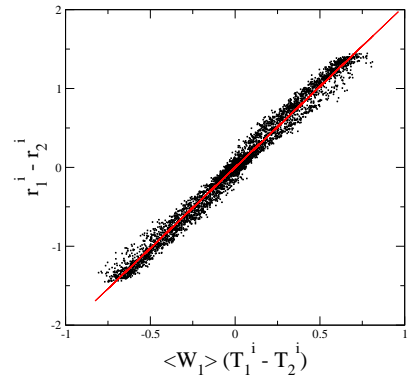,height=6cm}}} 
\caption{The variable 
$r_1^i-r_2^i$ versus $\langle W_1 \rangle (T_1^i - T_2^i)$. 
We find that $r_1^i-r_2^i \simeq \beta \langle W_1 \rangle (T_1^i -
T_2^i)$, with $\beta$ = 2.0512(3).}
\label{nova_fig11}
\end{figure}
\begin{figure}[!h]
\centerline{\hbox{\psfig{file=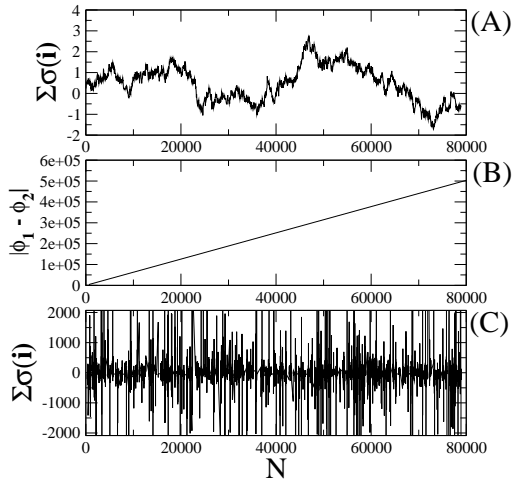,height=8cm}}} 
\caption{In (A), we show the quantity $\sigma$
in Eq. (\ref{novo_bound}) for a situation when PS exists.  As we
decrease the coupling, Eq. (\ref{PS}) is not anymore
satisfied as shown in (B),  as well as Eq. (\ref{novo_bound}),
 as shown in (C). In (C) we have made a zoom in of the
vertical axis. In (A), 
$\delta \alpha=0.001$ and $\epsilon$=0.01 and in (B-C), 
$\delta \alpha=0.001$ and $\epsilon$=0.000001.}
\label{nova_fig13}
\end{figure}
Then, we show in Fig.  \ref{nova_fig11} that the linear hypothesis between
$r_1^i-r_2^i$ and $\langle W_1 \rangle (T_1^i - T_2^i)$ done in Eq.
(\ref{relacao_fase_tempo}) stands and $\beta$=2.0512$\pm$0.0003. If PS is not
present, such linear scale is not anymore found for the system
considered.

In Fig. (\ref{nova_fig13})(A), we show the quantity $\sigma$
in Eq. (\ref{novo_bound}) for a situation that PS exists.  As we
decrease the coupling, Eq. (\ref{PS}) is not anymore satisfied as
shown in (B), as well as, Eq. (\ref{novo_bound}). In (C) we make
a zoom in of the vertical axis.  Note the different nature of the fluctuations of
the phase difference in (B) and the term $\sigma$ in Eq.
(\ref{novo_bound}).  That is because the term $\sigma$ represents the
phase difference without the linear growing trend, responsible to 
make the phase difference in (B) to present a positive slope. 

In order to compare the phase as defined in Eq.
(\ref{integral_vetor_tangente}) (for $\delta \alpha$=0.001 and
$\epsilon$=0.01), and the phase as defined in \cite{park}, e.g.
$\theta = tg(y/x)$, we compare the function $\langle r_j \rangle$, as
calculated for both definitions.  For the phase, as defined in Eq.
(\ref{integral_vetor_tangente}), we arrive at $\langle r_j
\rangle$=6.2984 and so, $\langle r_j \rangle > 2\pi$. Other quantities
are $\langle W_j \rangle$=0.1651, and $\langle T_j \rangle$ = 6.07097.
On the other hand, the phase as defined in \cite{park} is a function
that grows in average 2$\pi$ each time the trajectory crosses some
Poincar\'e section, which gives $\langle r_j \rangle= 2\pi$.  So, the
phase definitions arrive at two different quantities, but Eq.
(\ref{PS}) is valid in order to detect PS and Eq.
(\ref{relaciona_r_T}) is valid to measure the average angular
frequency of the attractor, using both phase definitions.

\begin{figure}[!h]
\centerline{\hbox{\psfig{file=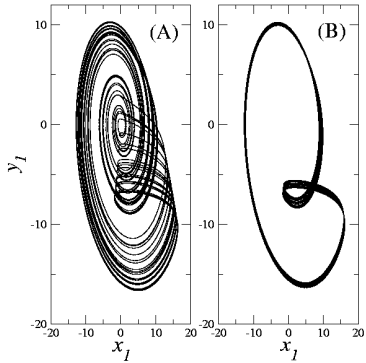,height=6cm}}}
\caption{Chaotic attractors projected on the variables $x_1$ and
  $y_1$.  (A) The coupling is null and therefore, there is no PS. Here
  one sees the non-coherent Fannel attractor.  (B) The coupling
  induces PS, creating a coherent dynamics.}
\label{coupled_fannel}
\end{figure}

\begin{figure}[!h]
\centerline{\hbox{\psfig{file=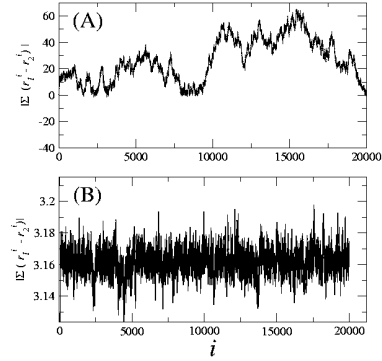,height=6cm}}}
\caption{Discrete phase difference $| \sum_{i=0}^{N-1} (r^i_1 -
  r^i_2)|$ for no coupling (A) where there is not PS and for a
  coupling $\epsilon$ = 0.00535 (B) responsible to induce PS.}
\label{PS_fannel}
\end{figure}

To illustrate the generality of the phase definition in Eq.
(\ref{integral_vetor_tangente}) in order to detect the phenomena of PS
in non-coherent attractors, we consider Eqs. (\ref{rossler}) with the
following set of parameters, $a$=0.3, $b$=0.4, $d$=7.5, such that we
have the fannel attractor shown in Fig. \ref{coupled_fannel}(A). This
attractor has a non-coherent phase character \cite{osipov,kreso}.  For
a parameter mismatch of $\delta \alpha=0.0002$, and for a null
coupling, $\epsilon$=0, both R\"ossler oscillators (presenting the
fannel attractor) are not phase synchronized as one can check in Fig.
\ref{PS_fannel}(A), which shows the absolute discrete phase difference
in Eq. (\ref{nova3}). As we introduce the coupling $\epsilon$ =
0.00535, the oscillators presents weak coherent motion.

\section{Phase synchronization in two coupled neurons}

\begin{figure}[!h]
\centerline{\hbox{\psfig{file=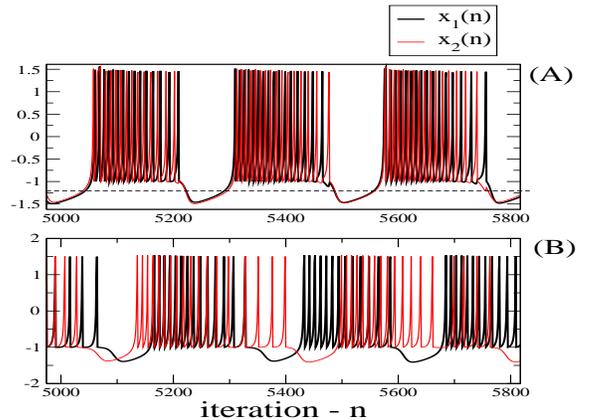,height=6cm,width=9cm}}}
\caption{A sample of the variables $x_1(n)$ and $x_2(n)$, from the subspaces that 
  correspond to both neurons, for a situation where there is PS, for
  $\epsilon$=0.03 (A), and for a situation where there is not PS, for
  $\epsilon$=0.001 (B). In (A), Eq.  (\ref{discrete_difference}) is
  satisfied, and in (B) is not. In (A), we show three bursts, which
  are basically a sequence of spiking.}
\label{nova_fig5}
\end{figure}

\begin{figure}[!h]
\centerline{\hbox{\psfig{file=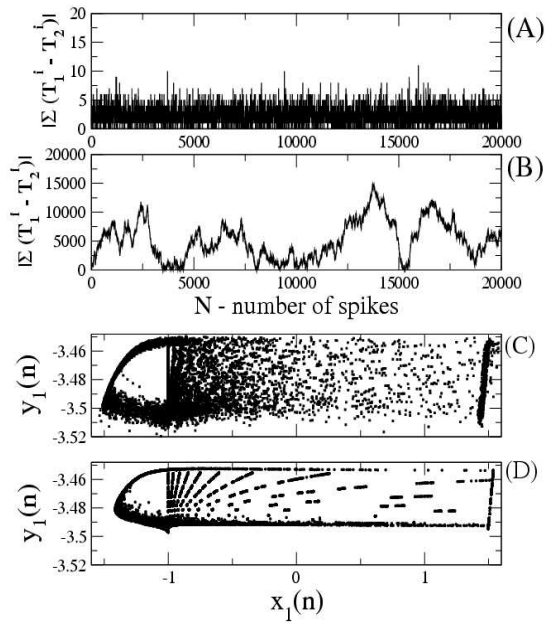,height=9cm,width=8cm}}}
\caption{In (A-B), we show the absolute difference between the time of
  the $N$-th burst, in both subspaces (that represent the two
  neurons).  In (A), Eq.  (\ref{discrete_difference}) is satisfied
  with $\langle T_1 \rangle$=259.028 and, in (B), Eq.
  (\ref{discrete_difference}) is not satisfied (there is not PS).  For
  (A) and (C), $\alpha_1$=4.99 and $\epsilon$=0.03.  For (B) and (D),
  $\alpha_1$=4.99 and $\epsilon$=0.001.}
\label{ergodic_potsdam_fig02}
\end{figure}

Now, we give an example for the detection of PS without the knowledge
of the state equations, but either only using a time series of bursting 
events. We consider two
non-identical coupled neurons described by 
{\small
\begin{eqnarray}
x_{j}({n+1})&=& f[x_j(n),y_j(n)+\beta_j(n)] \label{rulkov_map} \\
y_j({n+1})& =& y_j(n) - \theta(x_j({n})+1)+\theta\sigma_j+\theta\beta_j(n), 
\end{eqnarray}}
\noindent
which produces a chaotic attractor, for $\theta$ = $0.001$,
$\alpha_2$=5, $\sigma_1=0.240$, and $\sigma_2$=0.241.  The subspaces
are defined as $\mathcal{X}_j$=$({x_j,y_j})$.  $\beta_{1,2}(n)$ =
$g[x_{2,1}(n)-x_{1,2}(n)]$. The function
$f$ is given by {\small
\begin{eqnarray}
 f = \alpha_j/[1-x_j(n)]+y_j(n), \mbox{\ \ \ } x\leq 0 \nonumber \\
f = \alpha_j + y_j(n), \mbox{\ \ \ } 0<x<\alpha_j(n)+y \label{neuronio} \\
f = -1, \mbox{\ \ \ } x\geq \alpha+y_j(n) \nonumber 
\end{eqnarray} }
\noindent
The control parameters are $\alpha_1$ and $g$, with
$|\alpha_1-\alpha_2|$ being the parameter mismatch and $g$ the
coupling amplitude (cf. \cite{rulkov}).  The time at which events
occur  is defined by measuring the
time instants in which the variable $x_j$, of subsystem
$\mathcal{X}_j$, is equal to $x_j=-1.2$ (the event is the occurrence
of a burst) \cite{explica_threshold}, and $N_j$ is the number of
bursts of the subsystem $\mathcal{X}_j$. In this example, PS exists if
Eq.  (\ref{discrete_difference}) is satisfied, which also means that
Eq.  (\ref{event_synchrony}) is satisfied.

In Fig. \ref{nova_fig5}, we show the variables $x_1(n)$ and $x_2(n)$ for
a situation where there is PS (A), and for a situation where there is
not PS (B). Note that in (A), although the neurons are phase
synchronized, the difference between the number of bursts in the
variable $x_1(n)$ minus the number of bursts in the variable $x_2(n)$
might be different than zero (for a short moment), as the hypothesis done in Eq.
(\ref{event_synchrony}). In (A), we also represent by the dashed line
the threshold, $x_j=-1.2$, from which the events are specified.

In Fig. \ref{ergodic_potsdam_fig02}(A-B), we show the absolute
difference between the time of the $N$-th burst, in both neurons.  In
(A), Eq. (\ref{discrete_difference}) is satisfied (there is PS) with
$\langle T_1 \rangle$=259.028, and therefore 0.5$\langle T_1
\rangle$=124.5014, much bigger than the maximum fluctuation in (A). In
(B), there is no PS.  In (C) and (D), we show a projection of the
attractor on the variables $(x_1,y_1)$.  In (B) and (D), $\langle T_1
\rangle$ = 396.964, and $\langle T_2 \rangle$ = 398.407.

Note that although the attractor of these neurons have not the
dynamics of a limit cycle, presenting a very complicated geometry in
the phase space (as one can see in Fig. \ref{ergodic_potsdam_fig02}),
it is still possible to well define events as well as the average
period of the spiking times by the use of the threshold shown in Fig.
\ref{nova_fig5}, a characteristic that defines this attractor to be of
the weak coherent type.

\section{Conclusions}

We estimate the inferior bound value of the absolute phase
difference between two coupled chaotic systems, in order to verify the
existence of phase synchronization between them. Our approach shows that this bound
value $\langle r \rangle$ is given by the average evolution of the
phase, calculated in a subspace of the attractor, for a series of
pairs of events in this same subspace.  These events
can be the number of local maxima or minima in the trajectory, the
crossing of the trajectory to some Poincar\'e section, or the
occurrence of a burst/spike.

This result was achieve because we can inspect for phase
synchronization looking for the phase difference at the times for
which the same number of events happens in both subsystems.  The
advantage of looking at the phase difference at these times, instead
of looking at the continuous phase difference, is that this approach
allows us to detect phase synchronization by looking for a bounded
time difference between events. This is helpful for chaotic systems
from which there is no available information about the state
equations.

If only the number of events is available, one can also find evidences
of PS by checking that the absolute difference between the number of
events has to be smaller or equal than 1. This allows to infer the
existence of PS in maps. These maps can be derived either from a flow
(by measuring events as the trajectory crosses a Poincar\'e section,
or by detecting local maxima of the trajectory) or they can be
dynamical systems with a discrete formulation.  

In this work, the phase is introduced to be a quantity that measures
the amount of rotation of the tangent vector of the flow. 

All our results are extended to coupled chaotic systems that present
coherent properties as defined in \cite{kreso}, i.e., it is possible
to define an average time between two events $\langle T \rangle$, such
that for each returning time $\tau$, it is true that $| \tau - \langle
T \rangle |<\kappa$, with $\kappa \ll \langle T \rangle$.

\section{Acknowledgment} 

MSB would like to thank Alexander von Humboldt
Foundation. TP and JK would like to thank the Helmholz Center for
Mind and Brain Dynamics and SFB 555.

\appendix

\section{Contructing PS-sets from the event time series}\label{apendice1}

The event time series $\tau_j^i$ can be used to construct maps of the
attractor, whose geometrical properties states whether there is PS.
These maps are constructed following simple rules:
  
\begin{itemize}
\item At the time $\tau_2^i$, a point of the trajectory in $\mathcal{X}_1$ is collected.
                      
\item At the time $\tau_1^i$, a point of the trajectory in $\mathcal{X}_2$ is collected.
\end{itemize}

So, as a result of measuring the trajectories in $\mathcal{X}_1$
(resp.$\mathcal{X}_2$) at the times $\tau_2^i$ (resp $\tau_1^i$) we
have a discret set of points $\mathcal{D}_1$ ( resp.
$\mathcal{D}_2$).

In PS, these sets $\mathcal{D}_j$ will be localized, not spreading out
to the whole attractor. In this case $\mathcal{D}_j$ is called PS-set.
The theory for characthering and constructing these sets is presented
in \cite{baptista:2004}. In a short, what happens is the following: when phase
synchronization occurs, the times for a trajectory to pass through a
Poincare section (or reach some defined event) becomes relatively more
regular. Since we measure the trajectory on $\mathcal{X}_1$ by the
timing of events in $\mathcal{X}_2$, these maps are localized around
the Poincar\'e section.  For a non synchronous phase dynamics, the
sets $\mathcal{D}_j$ spreads over $\mathcal{X}_j$.  Thus, by detecting
a PS-set, one does not have to check if the inequality for the time
event difference holds.

In the following, we give examples of the PS-sets in the coupled
R\"ossler oscillators and in the Rulkov map, that mimics the neuronal
dynamics presenting spiking/burting behavior.

\subsection{PS-sets for the coupled R\"ossler oscillators}

The time series, $\mathcal{M}^{\tau}_j$, that define the events in
$\mathcal{X}_j$, are defined as follows.  $\mathcal{M}^{\tau}_1$: {\bf
  $\tau^i_1$} is constructed measuring the time the trajectory crosses
the plane $y_2$=0 in {\bf $\mathcal{X}_2$}; $\mathcal{M}^{\tau}_2$:
{\bf $\tau^i_2$} is constructed measuring the time the trajectory
crosses the plane $y_1$=0 in {\bf $\mathcal{X}_1$}.

\begin{figure}[!h]
 \centerline{\hbox{\psfig{file=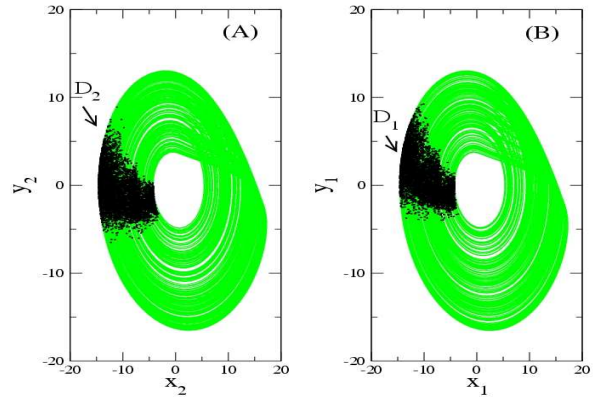,height=6cm,width=8cm}}}
\caption{Bidimensional projection of the attractor (gray) and 
  of the projections $\mathcal{D}_j$ of the PS-set (black) on the
  subspaces $\mathcal{X}_j$.  The PS-set projection
  $\mathcal{D}_2$, in (A), is constructed using time series
  $\mathcal{M}^{\tau}_2$, and $\mathcal{D}_1$, in (B), is constructed
  using time series $\mathcal{M}^{\tau}_1$.  $\delta \alpha=0.001$ and
  $\epsilon$=0.01.}
\label{ergodic_fig1}
\end{figure}

In Fig.  \ref{ergodic_fig1}, we show the coupled R\"ossler oscillators
for a situation where PS exists.  In this figure, we show
bidimensional projections on the variables of subsystem
$\mathcal{X}_2$ (A) and $\mathcal{X}_1$ (B). In gray, we show the
attractor projection, and in black, projections of the PS-set
$\mathcal{D}_1$ (A) and $\mathcal{D}_2$ (B).  Note that the PS-sets,
do not visit everywhere $\mathcal{X}_j$, rather are localized
structures. 

\subsection{PS-sets for the coupled Rulkov Map}

In the neuronal dynamics is not possible to define a Poincar\'e
section, due to the non-coherence of the attractor. However, it is
possible to define an event where the dynamics is weak coherent. This
event is the ending or the beggining of the bursts, and in here we
choose the beggining of the burst.  Hence, we construct our time series
by measuring the crossing of the trajectory with the threshold,
$x_j=-1.2$.

\begin{figure}[!h]
\centerline{\hbox{\psfig{file=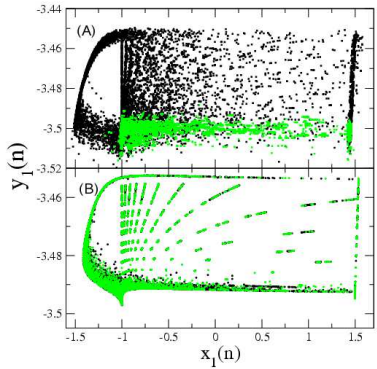,height=6cm}}}
\caption{In (A-B), we show the absolute difference between the time of the
  $N$-th burst, in both subspaces (that represent the two neurons).
  In (A), Eq.  (\ref{discrete_difference}) is satisfied (there is PS),
  with $\langle T_1 \rangle$=259.028 and, in (B), Eq.
  (\ref{discrete_difference}) is not satisfied (there is not PS).  For
  the same parameters as in (A), in (C) we show the PS-set
  projection $\mathcal{D}_1$ (gray), while for the same parameters as in (B), in
  (D) there is not a PS-set.  For (A) and (C), $\alpha_1$=4.99 and
  $\epsilon$=0.03.  For (B) and (D), $\alpha_1$=4.99 and
  $\epsilon$=0.001.}
\label{ps-neuron}
\end{figure}

In Fig. \ref{ps-neuron} we show a projection of the attractor on the
variables $(x_1,y_1)$, in black points, and the subsets
$\mathcal{D}_j$, in gray points.  In (A), where we have phase
synchronization the set $\mathcal{D}_1$ does not fullfil the whole
attractor, but is rather localized, whereas in (B), where PS is not
present the set $\mathcal{D}_2$ spreads over the attractor
$\mathcal{X}_2$.

\section{$\beta$ digression}\label{apendice2}

In this section we explain why in coherent attractors, e.g.
R\"ossler-type, the constant $\beta$ is approximately 2.

\begin{figure}[!h]
  \centerline{\hbox{\psfig{file=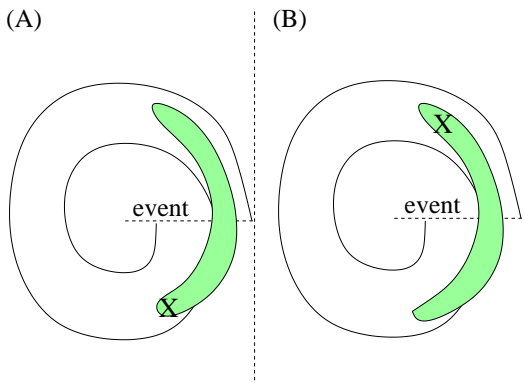,height=5.cm}}}
  \caption{Pictorial visualization of a situation where PS exists. We
    represent in (A) the trajectory in the subspace $\mathcal{X}_1$,
    and (B) the trajectory in the subspace $\mathcal{X}_2$.  The
    filled regions in (A) [(B)] represent trajectory positions at the
    time the $N$-th event happens in $\mathcal{X}_2$ [$\mathcal{X}_1$]
    for a situation when PS exists.  An event is considered to happen
    when the trajectory crosses the dotted line.}
\label{nova_fig10}
\end{figure}

That is so, because we compare the phase difference at the time events
occurrence. Let us just remember that we are measuring the phase in
one subsystem at the times that events in the other subsystem happen.
Hence, at the time events happens in $\mathcal{X}_{1}$ [resp.
$\mathcal{X}_{2}$], we collect points in $\mathcal{X}_{2}$ [resp.
$\mathcal{X}_{1}$], obtaining the gray filled region in Fig.
\ref{nova_fig10}(B) [resp. (A)], which is a PS-set.

In particular, when the $N$-th event happens in $\mathcal{X}_{2}$, the
trajectory on $\mathcal{X}_{1}$ is indicated by the cross in (A). At
this time, $\tau_2^N$, we record the phase in $\mathcal{X}_{1}$,
namely $\phi_1(\tau_2^N)$. As the time goes on, the trajectory (in a
unclockwise direction of rotation) on $\mathcal{X}_{1}$ reaches the
event line in $\mathcal{X}_{2}$ at the time $\tau_1^N$. At this time,
the trajectory in $\mathcal{X}_{2}$ is at the cross in (B), and the
phase is $\phi_2(\tau_1^N)$. Since these are typical events, we can
say that $|\tau_2^N - \tau_1^N| \approx \langle T \rangle/4$, for the
particular case represented in this figure. That is so because the
time difference is approaximately given by the time that the
trajectory in $\mathcal{X}_{1}$ spends from the cross in (A) till the
event line, which is approximately 1/4 of the average period $\langle
T \rangle$.

The phase difference, at which the same number $N$ of events happen,
is $|\phi_1(\tau_1^N) - \phi_1(\tau_2^N)| \approx \langle r
\rangle/2$, since this phase difference is basically given by the
displacement of the phase in $\mathcal{X}_1$ from the cross in (A)
till the event line, plus, the displacement of the phase in
$\mathcal{X}_2$ from the event line till the cross in (B). But that is
approximately given by 1/2 of the average increasing of the phase
$\langle r \rangle$, which was shown to be equal to $\langle W \rangle
\times \langle T \rangle$. Therefore, $|\phi_1(\tau_1^N) -
\phi_1(\tau_2^N)| \approx 2 \times |tau_2^N - \tau_1^N|$, which
consequently give us that $\beta \approx 2$.


\begin{thebibliography}{99}
\bibitem{cita_r} $r$ is considered to be a rational.
  However, as shown in Ref. \cite{murilo}, PS, as defined by the
  boundness of the phase difference was found in two chaotic systems
  for a finite but very large time interval, as $r$ approaches an
  irrational.  Therefore, although in this work we consider $r$ to be
  rational, we should make the remark that for the especial situation
  as the one presented in Ref.  \cite{murilo}, Eq. (\ref{PS}) can only
  be satisfied for a finite but large time if $r$ is considered to be
  irrational.
\bibitem{equilibrium} In Eq. (\ref{ps_cond_I}), we assume that the
  coupled system reached equilibrium, i.e., the asymptotic dynamics.
  Otherwise, one could introduce another finite constant inside the
  left side of this equation.
\bibitem{phase_definitions} Phase can be defined as a Hilbert
  transformation of a trajectory component, as the angle of a
  trajectory point into a special projection of the attractor, or as a
  function that grows $2\pi$, every time the chaotic trajectory
  crosses some specific surface.  In this work, we define phase to be
  a measure of the absolute rotation of the tangent vector in the
  subspaces $\mathcal{X}_j$.
\bibitem{calcula_analitico} A trivial case, where
  $|\dot{\vec{\mathcal{A}}}|$ can be analytically computed is in a
  planar circular motion. The position vector is $[x,y]=[\cos{(w t)},
  \sin(w t)]$, and the unitary tangent vector has the coordinates $[-
  \sin{(w t)}, \cos(w t)]$. The derivative of the tangent vector has
  the coordinates $\dot{\vec{\mathcal{A}}} = [- w \cos{(w t)}, -w
  \sin{w t}]$, and therefore, $|\dot{\vec{\mathcal{A}}}|$=$w$, and
  $\phi(t)$ = $wt$.
\bibitem{explica_threshold} This threshold enables to detect the time
  at which a burst starts.  In fact, due to small fluctuations of the
  trajectory close to the used threshold, $x_j$=-1.2, we only consider
  that a burst starts (the event happened), if after the trajectory
  having crossed the threshold at $x_j$=-1.2, it crosses eventually
  another threshold, positioned in $x_j$=0, before it crosses again
  the threshold $x_j$=-1.2, in the next event.
\end{thebibliography}
\end{document}